\documentstyle[prl,aps,multicol,epsf]{revtex}
\begin{document}
\draft

\title{Metal-Insulator Transitions in Interacting Disordered Systems}
\author{Myriam P. Sarachik}
\address{Physics Department, City College of the City University of 
New York, New York, New York 10031}

\maketitle

\vskip20mm

Metal-insulator transitions that occur in the limit of zero temperature in a 
variety of electronically disordered solids as a function of composition, dopant 
concentration, magnetic field, stress, or some other tuning parameter, have been the focus 
of study for many decades and continue to be a central problem in condensed 
matter physics.  A full overview of so broad a field would require considerably 
more space than I have been allotted.  I will thus limit my presentation 
to two specific areas that have been of particular interest to me.  
In the first part I will give a brief overview of the recent history and current 
status of the so-called ``critical exponent puzzle'' in three-dimensional 
doped semiconductors and amorphous metal-semiconductor mixtures, including 
some interesting recent developments.  The second part will be a brief 
summary of new and exciting findings in dilute 2D systems, such as silicon 
MOSFETs and GaAs/AlGaAs heterostructures, where the resistivity exhibits 
a metallic temperature 
dependence in some ranges of electron (hole) densities, raising the 
possibility of an unexpected metal-insulator transition in two dimensions.  
An equally intriguing property of these strongly interacting, low density 2D 
systems is their 
enormous magnetoresistance: for magnetic fields applied parallel to the electron 
plane, the resistivity increases dramatically by several 
orders of magnitude in response to relatively modest fields on the order 
of a few Tesla, saturating to a constant value at higher fields.

\section{The Critical Exponent Puzzle.}
Based on the Ioffe-Regel criterion, which asserts that the mean free path of an 
electron in a metallic system cannot be shorter than its wavelength, Mott\cite{mott} 
proposed in 1972 that there exists a minimum value of the conductivity, 
$\sigma_{min} = Ce^2/ha$, below which a system cannot be in the metallic phase; here 
$e$ is the electron charge, $h$ is Planck's constant, $a$ is the average distance 
between electrons at the critical concentration $n_c$, and $C$ was estimated by 
Mott to be a number between $0.15$ and $0.3$.  The conductivity is thus expected to 
drop discontinuously to zero at a Mott transition driven by 
interactions.  Some years later, the scaling theory of Abrahams, Anderson, 
Licciardello and Ramakrishnan\cite{gang} for disordered systems of 
non-interacting electrons implied that the transition is instead a continuous 
one, so that arbitrarily small values of conductivity are possible in the 
metallic phase. Theory soon followed which showed that the electrons (or holes) 
generally localize even more strongly when weak interactions are included 
\cite{leerama}.  Experiments in amorphous 
metal-semiconductor mixtures such as AuGe\cite{AuGe} and NbSi\cite{NbSi}, 
as well as in some doped semiconductors, most notably the elegant 
stress-tuning measurements by Paalanen {\it et al.} in Si:P\cite{paalanen}, 
provided strong evidence that the transition is indeed continuous.

Following 
these early seminal contributions, however, the field has been plagued for 
nearly two decades by controversy and conflicting results.  On the one hand, 
a consensus emerged rather quickly regarding the behavior of the amorphous 
metal-semiconductor mixtures: tunneling experiments\cite{AuGe,NbSi} 
documented the appearance 
of a square-root singularity in the density of states as the transition is 
approached from the metallic side, consistent with theoretical predictions for 
interacting electrons\cite{leerama}; the critical exponent $\mu$ 
which characterizes the (continuous) approach of the conductivity to the 
transition, $\sigma \approx (n-n_c)^\mu$, was found to be close to $1$ in 
essentially all the amorphous systems examined.  In contrast, conflicting results 
have been reported in doped semiconductors, and these materials continue to be the 
subject of debate and uncertainty as discussed in more detail below.  It 
should be noted, however, that a number of investigators, most actively 
A. Moebius\cite{moebius}, continue to question the generally accepted view; 
based on a scaling analysis of the conductivity of 
amorphous CrSe, this school maintains that the transition is discontinuous with a 
minimum metallic conductivity, 
as originally postulated by Mott.

	Electron interactions are known to play a role in the case of doped 
semiconductors.  This was demonstrated by tunneling experiments in 
Si:B\cite{marklee} 
which yielded results very similar to those found for GeAu and NbSi: a 
square-root singularity develops as the transition is approached from the 
metallic side, and a ``Coulomb gap'' appears in the insulator as the dopant 
concentration is further decreased.  However, a great deal of disagreement 
persists concerning the critical behavior of the conductivity, with various 
laboratories reporting different results on different, and sometimes even the same, 
semiconductor systems.

\vbox{
\vspace{0.2in}
\hbox{
\hspace{-0.2in} 
\epsfxsize 7in \epsfbox{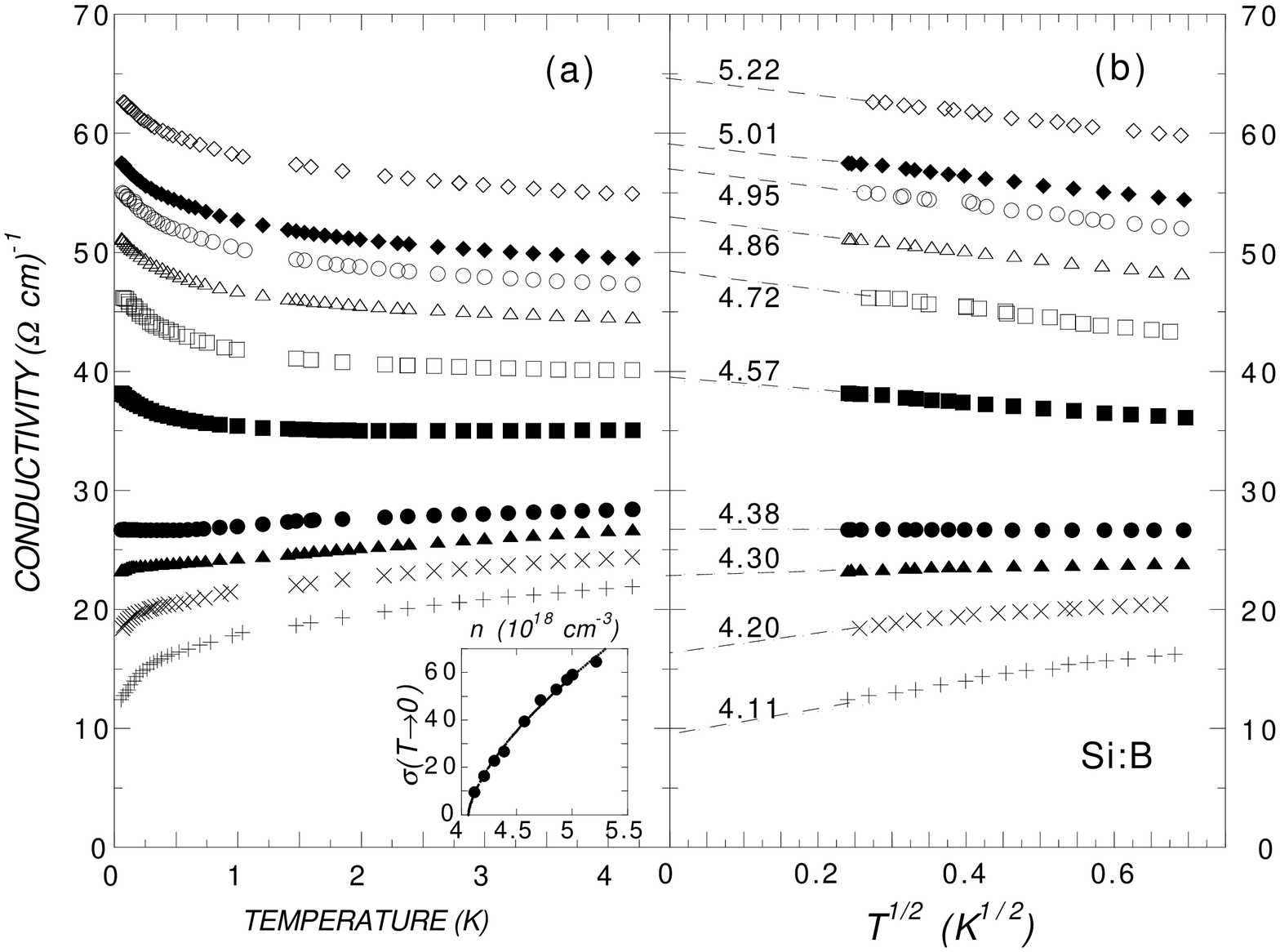} 
}
}
\refstepcounter{figure}
\parbox[b]{7in}{\baselineskip=12pt FIG.~\thefigure.
Conductivity of Si:B plotted as a function of: (a) temperature $T$; and (b) 
$T^{1/2}$.  Each curve corresponds to a different dopant concentration, 
labelled in units of $10^{18}$ cm$^{-3}$.  The inset shows zero-temperature 
extrapolations plotted as a function of dopant concentration.
\vspace{0.3in}
}
\label{1}

Typical curves are shown in Fig. 1, where the 
conductivity of Si:B is plotted as a function of temperature $T$ in frame (a), and 
as a function of $T^{1/2}$ in frame (b).  The conductivity is 
given by\cite{leerama}: 
$$
\sigma(T)= \sigma(0) +\Delta \sigma_{int}+\Delta \sigma_{loc}= 
\sigma(0) + mT^{1/2} + BT^{p/2}
$$
The second term on the right-hand side is due to electron-electron interactions 
and the last term is the correction to the zero-temperature 
conductivity due to localization.  
The exponent $p$ reflects the temperature dependence of the scattering rate, 
$\tau_\phi^{-1} \propto T^p$ of the dominant phase-breaking mechanism 
responsible for delocalization, such as electron-phonon scattering or 
spin-orbit scattering.  The last term is assumed small at very low 
temperatures, and the conductivity is generally plotted as a function of 
$T^{1/2}$, as in Fig. 1 (b).  There is little theoretical justification for 
using this expression within the critical range, where it is applied ``by 
default'' in a region where the behavior of the conductivity is not known.  
We note that the slope $m$ that determines the 
conductivity at low temperatures changes sign from negative to positive as 
the metal-insulator transition is approached.  The significance of this change 
of sign has been a matter of debate, and lies at the heart of the controversy 
regarding the critical behavior.  

The critical conductivity exponent $\mu$ is generally determined by the following 
procedure.  For each 
sample with a given concentration (or for each value of stress, magnetic field, or 
other tuning parameter), a single, zero-temperature extrapolated value of the 
conductivity is deduced from data obtained at finite temperatures.  The 
zero-temperature extrapolations are then plotted as a function of concentration 
(or stress, or field) to obtain the critical behavior, as indicated in the 
inset to Fig. 1 (a).  This experimental protocol has yielded conflicting values 
for $\mu$.  Important examples are shown in the 
next two figures.

\vbox{
\vspace{0.2in}
\hbox{
\hspace{1.5in} 
\epsfxsize 4in \epsfbox{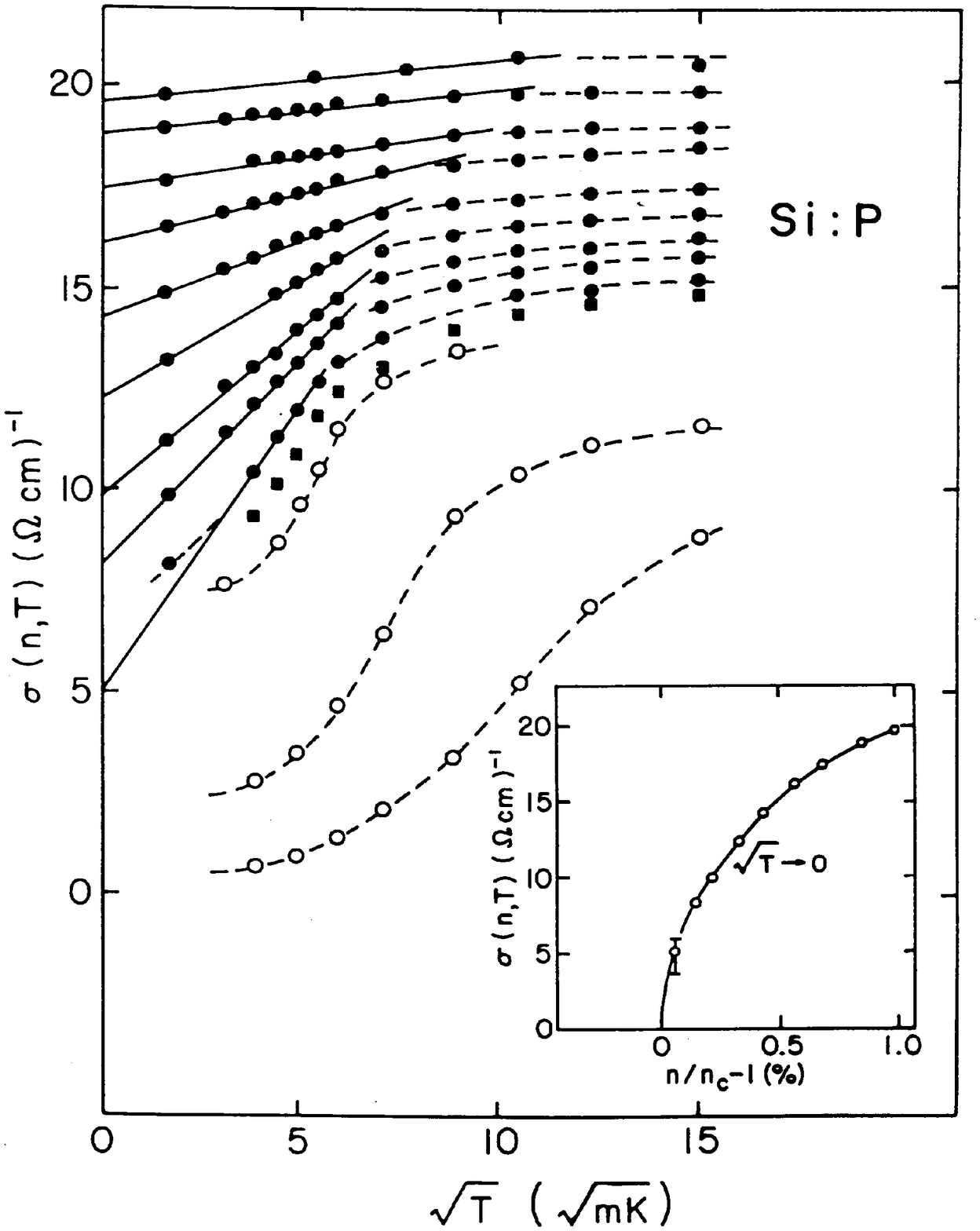} 
}
}
\refstepcounter{figure}
\parbox[b]{7in}{\baselineskip=12pt FIG.~\thefigure.
Data of Paalanen {\it et al.} (ref. 6) for the conductivity as a function 
of temperature of uniaxially stressed Si:P very near the metal-insulator 
transition; the inset shows zero-temperature extrapolations plotted as a function 
of stress, yielding an exponent $\mu=1/2$.
\vspace{0.3in}
}
\label{2}

Figure 2 shows early classic measurements of the conductivity of uniaxially-stressed 
Si:P taken down to unusually low temperatures by the Bell group\cite{paalanen}; 
fitted to the square-root temperature dependence of Eq. (1), the zero-temperature 
extrapolations yielded a critical exponent $\mu =1/2$ (see the inset). If the 
correlation length exponent $\nu$ is equal to the 
conductivity exponent $\mu$, as expected within Wegner scaling\cite{wegner}, 
this violates a lower bound $\nu>2/3$ in three dimensions calculated by 
Chayes {\it et al.}\cite{chayes}.  A possible solution to the exponent puzzle was 
subsequently proposed by the Karlsruhe group of H. v. Lohneysen\cite{stupp} based on 
measurements in unstressed Si:P shown in Fig. 3.  Stupp {\it et al.} suggested that 
only those samples for which the low temperature slopes of the conductivity are 
positive are in the critical region and should be used to deduce the critical 
behavior.  As shown in Fig. 3 (b), the Karlsruhe experiments yielded a much larger 
exponent, $\mu \approx 1.3$, based on a restricted range of dopant concentrations 
very near a critical concentration that is assumed to be substantially smaller than 
the generally accepted value.  The critical exponent near $1/2$ obtained by the Bell 
group was attributed by these authors to the improper inclusion of samples outside 
the critical region.  
The Bell group strongly disputed this claim, and attributed the large Karlsruhe 
exponent to inhomogeneities that cause ``rounding'' near the transition
\cite{comment,numbers}.  
Indeed, the unknown breadth of the critical region 
had been a source of some concern, and the Karlsruhe ansatz offered a 
relatively simple and attractive solution to a vexing problem.  At the same 
time, however, carefully executed investigations by Itoh {\it et al.}\cite{itoh1} 
of the conductivity of neutron transmutation doped Ge:Ga on both sides of the 
metal-insulator transition provided strong

\vbox{
\vspace{0.2
in}
\hbox{
\hspace{1.5in} 
\epsfxsize 4in \epsfbox{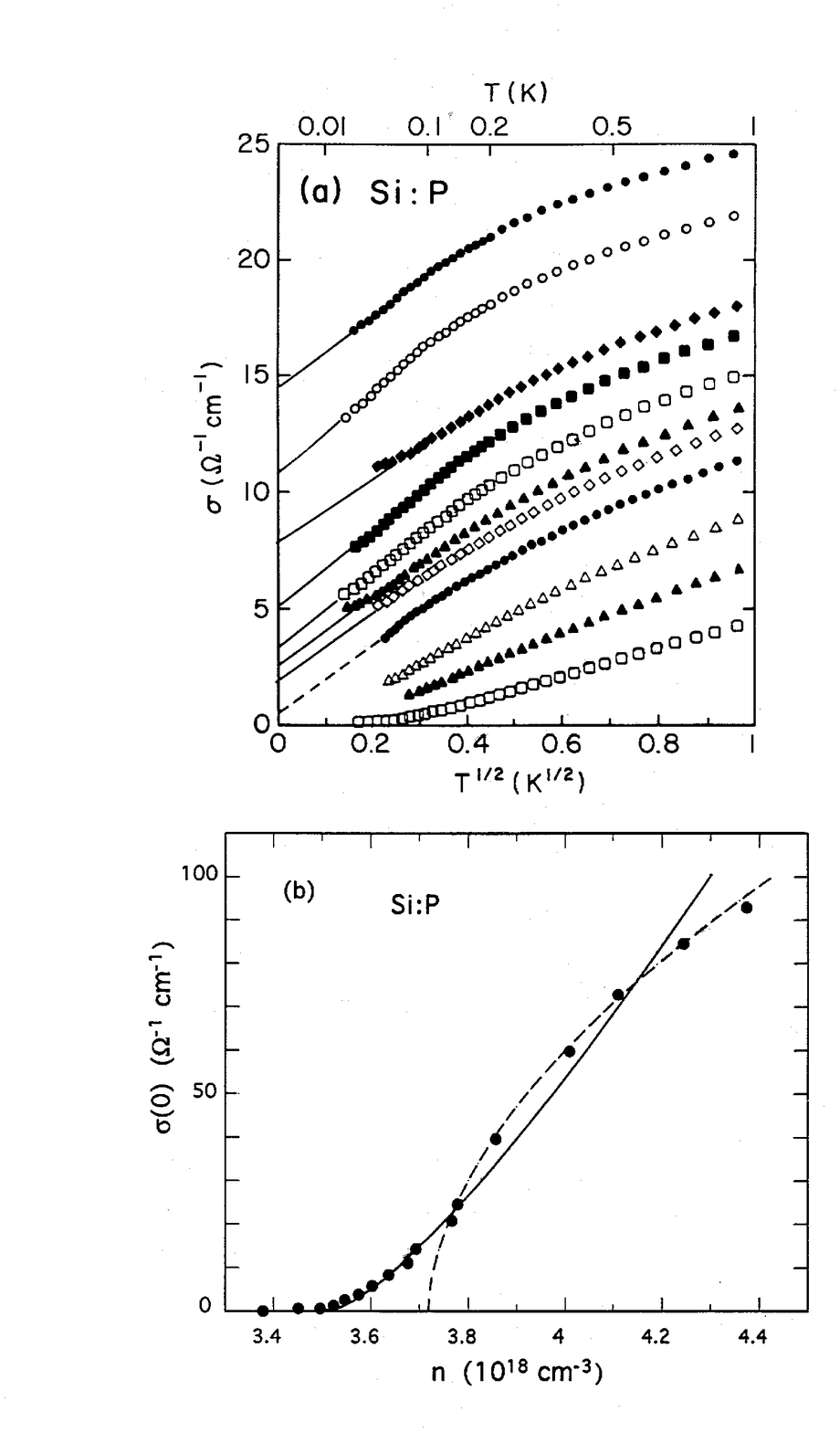} 
}
}
\refstepcounter{figure}
\parbox[b]{7in}{\baselineskip=12pt FIG.~\thefigure.
Data of Stupp {\it et al.} (ref. 11) for the conductivity as a 
function of 
temperature of a series Si:P samples with closely spaced dopant concentrations 
very near the metal-insulator transition.  (b)  The conductivity extrapolated to 
zero-temperature as a function of dopant concentration.  
Stupp et al. (1993) claim that restricting the analysis to samples whose 
resistivities have positive slopes at low temperatures yields a true critical 
conductivity exponent $\mu\approx 1.3$ (solid line), while including points further 
from the 
transition (claimed to be outside the critical region) results in an apparent 
exponent $\mu\approx 1/2$ (dashed line).
\vspace{0.3in}
}
\label{3}
evidence for the smaller exponent around $1/2$.  To 
complicate matters further, Castner\cite{castner} contended that some of the 
positive-slope curves that were classified as metallic actually obey Mott 
variable-range hopping, placing these on the {\it insulating} side of the 
transition.  The few 
who were directly involved in the issue divided 
rather sharply into opposing camps, while the rest of the community began to 
lose interest in a problem that was making little apparent headway.

It has been difficult to obtain reliable determinations of the critical 
behavior for a number of reasons.  One important issue is whether the 
distribution of dopant atoms is statistically random.  This problem can be 
minimized by doping through neutron transmutation, as has been done in 
Ge:Ga\cite{itoh1}.  Another 
difficulty is that the approach to the transition is most often controlled 
experimentally by varying the dopant concentration, $n$, near its critical 
value, $n_c$, a method that entails the use of a discrete set of carefully 
characterized samples.  This makes it difficult to do systematic, controlled 
studies on closely spaced samples very near the transition within the critical 
regime.  This problem has been circumvented in a few studies where individual 
samples have been driven through the transition using a different tuning 
parameter such as uniaxial stress or magnetic field.  The central problem, 
however, is that zero-temperature conductivities deduced from extrapolations 
from finite temperature measurements are uncertain and unreliable, 
particularly in the absence of any theory known to be valid in the critical 
region.

\vbox{
\vspace{0.2
in}
\hbox{
\hspace{1.5in} 
\epsfxsize 4in \epsfbox{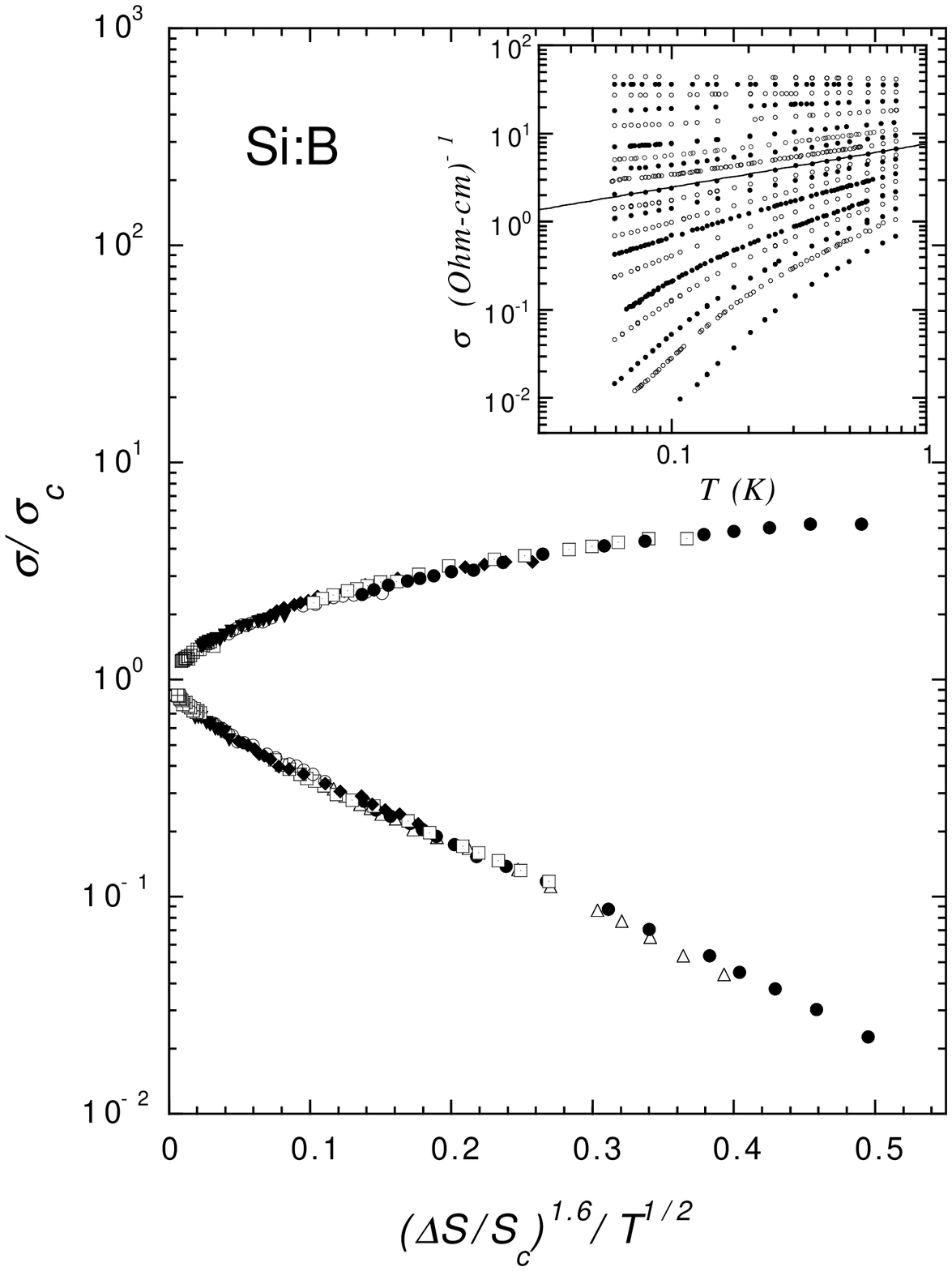} 
}
}
\refstepcounter{figure}
\parbox[b]{7in}{\baselineskip=12pt FIG.~\thefigure.
For different values of the tuning parameter, $S$, the normalized 
conductivity, $\sigma/\sigma_c$, of uniaxially stressed samples of Si:B is shown 
on a log-log 
scale as 
a function of the scaling variable $[(\Delta S)/S_c]/T^{1/z\nu}$, with $z\nu = 3.2$.  
Here $\Delta S = (S - S_c)$, where $S_c$ is the critical stress; the critical 
temperature dependence at the transition is $S_c  \propto T^{1/2}$.  The inset 
shows the unscaled conductivity as a function of temperature on a log-log scale 
for different values of stress.
\vspace{0.3in}
}
\label{4}

\vbox{
\vspace{0.2
in}
\hbox{
\hspace{1.5in} 
\epsfxsize 4in \epsfbox{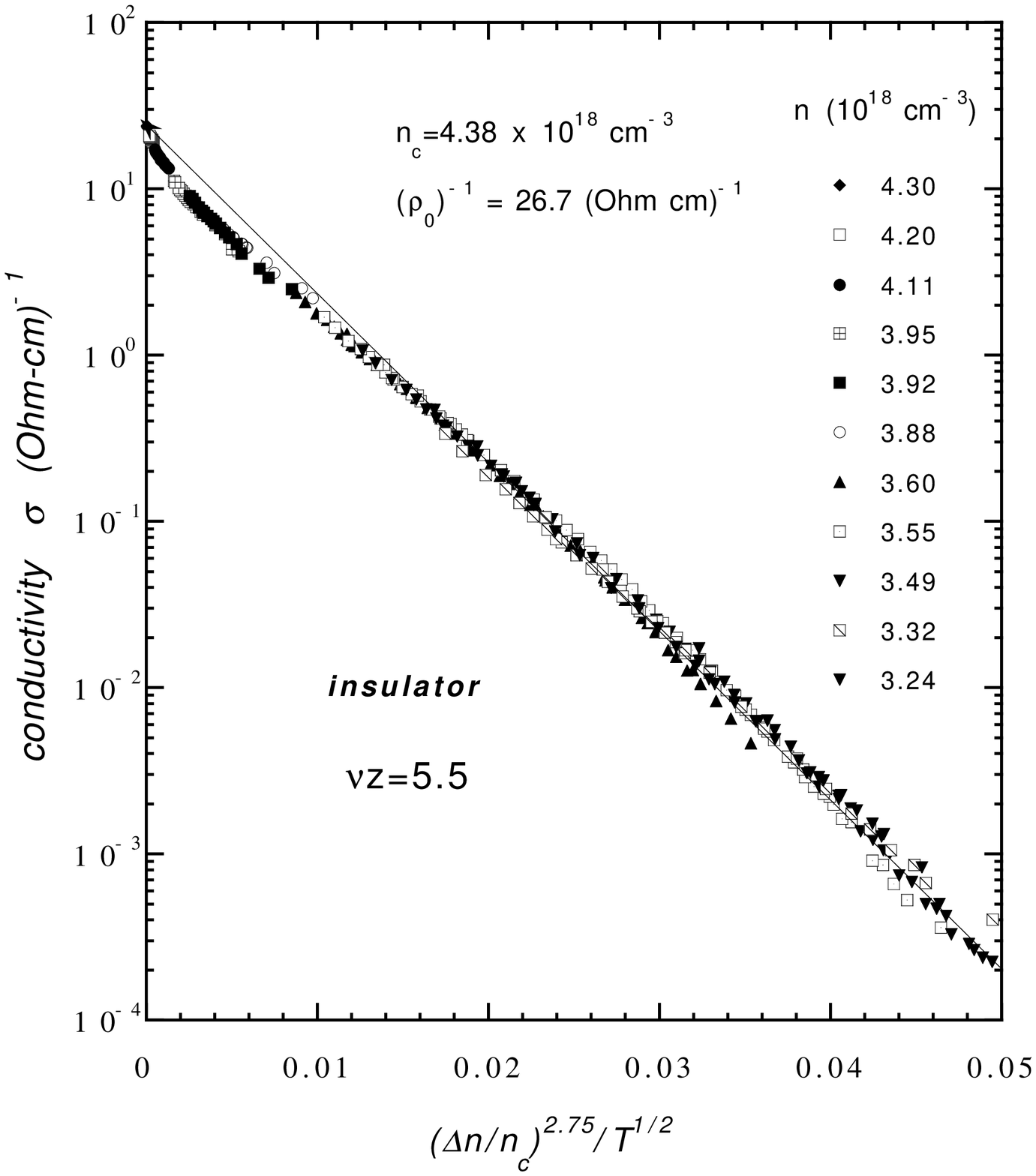} 
}
}
\refstepcounter{figure}
\parbox[b]{7in}{\baselineskip=12pt FIG.~\thefigure.
For different values of dopant concentration $n$ used here as the 
tuning parameter, the conductivity  
$\sigma$ of unstressed Si:B is shown on a logarithmic scale 
as a function of $(T^*/T)^{1/2}$; here $T^*\propto (\Delta n)^{z\nu}$.
\vspace{0.3in}
}
\label{5}

A full scaling analysis which uses data obtained at all temperatures obviates 
the need for extrapolations to zero\cite{moebius,belitzreview}.  
Attempts to apply finite-temperature scaling to the conductivity of crystalline doped 
semiconductors had largely failed until quite recently (except in the presence 
of an externally applied magnetic field).  Application by Belitz and Kirkpatrick
\cite{belitz} of finite temperature scaling to data for Si:P gave satisfactory 
results only over a severely restricted range of temperature, and yielded critical 
conductivity exponents $\mu = 0.29$ for the Bell data, and $\mu = 1$ for the 
Karlsruhe data\cite{numbers}.  No data were available from either group for the insulating 
side of the transition; as shown below, the availability of data on both sides 
of the transition imposes important constraints.

Potentially important progress on this question was recently achieved in 
experiments on stress-tuned Si:B\cite{bogdanovich}.  As shown in Fig. 4, full 
scaling of the conductivity with temperature and stress of the 
form\cite{belitzreview}:
$$
\sigma(n,T) = \sigma (n_c,T) F(\Delta n/T^{1/z\nu}) 
$$
was demonstrated on both sides of the transition.  Here 
$\sigma(n_c) \propto T^{\mu/z\nu}$, $n_c$ is the critical concentration, 
$\Delta n = (n - n_c)$, $\nu$ is the critical exponent that characterizes the 
divergence of the length 
scale, and $z$ and $\mu$ are the dynamical exponent and critical conductivity 
exponent, respectively.  On a log-log scale, the inset to Fig. 4 shows the 
(unscaled) conductivity for various values of stress.  The critical curve is denoted 
by the straight corresponding to a power law; the temperature dependence at the 
critical point is found to be $\sigma \propto T^{1/2}$ in stressed Si:B.  It is worth 
emphasizing again that the power of this method lies with the fact that {\it all} 
the data obtained at all temperatures are used in the scaled curves of Fig. 4, 
rather than a single zero-temperature extrapolation for each stress deduced from a 
full curve of the conductivity as a function of temperature.  
This imposes more 
stringent constraints, particularly when data are available on both sides of the 
transition, and yields critical exponents that are far more 
reliable and robust.

It is puzzling that the experiment on stress-tuned Si:B yielded a  
critical conductivity exponent $\mu=1.6$ considerably larger than any 
previous determination.  Instead of answering old questions, this raises new ones.  
For example, does 
the use of stress as a tuning parameter 
yield the same physics as varying the concentration, as had always 
been assumed?   Although it has not resolved the controversy regarding the value 
of $\mu$, the Bogdanovich {\it et al.} experiment demonstrated that finite 
temperature-scaling can be applied.  This has triggered a number of new attempts to 
use the same method in other cases.  Among these are reports from Karlsruhe
\cite{karlsruhestress} 
of scaling for stressed Si:P (yielding $\mu=1$), and 
by Itoh's group in Ge:Ga\cite{itohhamburg}.

I will close this section by showing some surprising results of a reanalysis of old 
data taken at City College on unstressed samples of Si:B.  Encouraged by the full 
scaling form that was successcully applied to stressed Si:B, an equivalent analysis 
was attempted for data obtained earlier by Peihua Dai where dopant concentration 
rather than stress was used to tune through 
the transition.  As mentioned earlier, in the case of stressed Si:B the critical 
curve exhibits a power-law dependence on temperature, 
$\sigma\propto T^{1/2}$; more generally, the critical temperature dependence of the 
conductivity has been reported in various 
different semiconductor systems as either $T^{1/2}$ or $T^{1/3}$.  Although the 
data can be manipulated to yield scaled curves for either the metallic or the 
insulating branches, neither of these choices for the critical T-dependence 
yields scaling on {\it both} sides of the transition for the conductivity of 
unstressed Si:B.  Surprisingly, the full scaling that was obtained for the stressed 
samples appears not to hold for the unstressed case.  On the other hand, as shown in 
Fig. 5, all the data obtained for insulating samples down to $0.75 n_c$ (which is 
clearly outside the critical region) collapse onto a single curve if 
one chooses to plot $\sigma$ itself rather than $\sigma/T^x$ with 
$x=\mu/z\nu=1/2$ or $1/3$.  The conductivity for concentrations ranging from 
$0.75 n_c$ to $n_c = 4.38 \times 10^{18}$ cm$^{-3}$ collapses onto a single curve; 
although three positive-slope samples (the lowest three curves of Fig. 1) are 
included that have generally been assumed to be on the metallic side of the 
transition, further careful work is required to determine whether the 
collapse holds or whether it breaks down very near the transition.  The conductivity 
of unstressed Si:B on the insulating side of the transition is thus given by 
$\rho=\rho_0 F(T^*(n)/T)$ with $T^* \propto (n_c - n)^y$, $F(0) = 1$, and a 
prefactor $\rho_0$ that is independent of temperature $T$ and dopant concentration 
$n$; this implies a resistivity $\rho = \rho_0$ that is independent of temperature 
at the critical point $n=n_c$.  Deep in the insulating phase the conductivity obeys 
exponentially activated Efros-Shklovskii variable-range hopping, 
$\rho = \rho_0$ exp$[-(T^*/T)^{1/2}$.

It is remarkable that, despite considerable effort over a period of decades, a 
complete and satisfactory understanding of the behavior of doped semiconductors 
near the metal-insulator transition has not yet emerged; this remains 
one of the most interesting and important open questions in condensed matter physics.

\section{Novel phenomena in dilute 2D systems: New Physics or Old?}

While we continue our efforts to understand the metal-insulator transition in 
three dimensions, new developments in disordered, dilute 
two-dimensional systems have opened an entirely new area of exciting physics, 
launched by the availability of silicon MOSFETs with unusually high mobilities 
fabricated in the (former) Soviet Union.  These samples allowed measurements at 
substantially lower densities than had been accessible earlier, a regime where 
interaction energies are very large compared with the Fermi energy.  The 
experimental findings have called into question our long-held belief
\cite{gang,leerama} that there is no metallic phase in two dimensions in the 
limit $T \rightarrow 0$.  The strange and enigmatic behavior of dilute 2D 
systems\cite{2Dreview} is illustrated in the next two figures.

The resistivity of a high-mobility silicon metal-oxide-semiconductor 
field-effect transistor (MOSFET) is shown in Fig. 6 (a) as a function of density 
at different temperatures, and in Fig. 6 (b) as a function of temperature at 
different densities\cite{kravchenko}.  The crossing point in frame (a) indicates a 
critical density $n_c$ below which the behavior is insulating, and above which the 
resistivity decreases with decreasing temperature, behavior that is normally 
associated with a metal (see frame (b)).  The curves shown in Fig. 1 (b) as a 
function of temperature $T$ can be collapsed onto two branches 
by applying a single scaling parameter $T_o$, a feature generally associated with 
quantum phase transitions.  These claims first met with considerable skepticism, 
but were soon confirmed in MOSFETs obtained from different sources, and similar 
behavior was subsequently found for other two-dimensional systems 
(p-GaAs, n-GaAs, p-SiGe, etc.).

As is true for many other interesting open questions in condensed matter physics, 
both interactions and disorder play a role, and their relative importance is 
unclear: (i) the transition from insulating to metallic temperature-dependence 
occurs at very low electron (hole) densities ($\approx 10^{11}$ cm$^{-2}$ or lower), 
where interaction energies are much larger than kinetic energies; (ii) 
the resistivity is on the order of $h/e^2$, suggesting that disorder plays a role.

\vbox{
\vspace{0.2
in}
\hbox{
\hspace{1in} 
\epsfxsize 5in \epsfbox{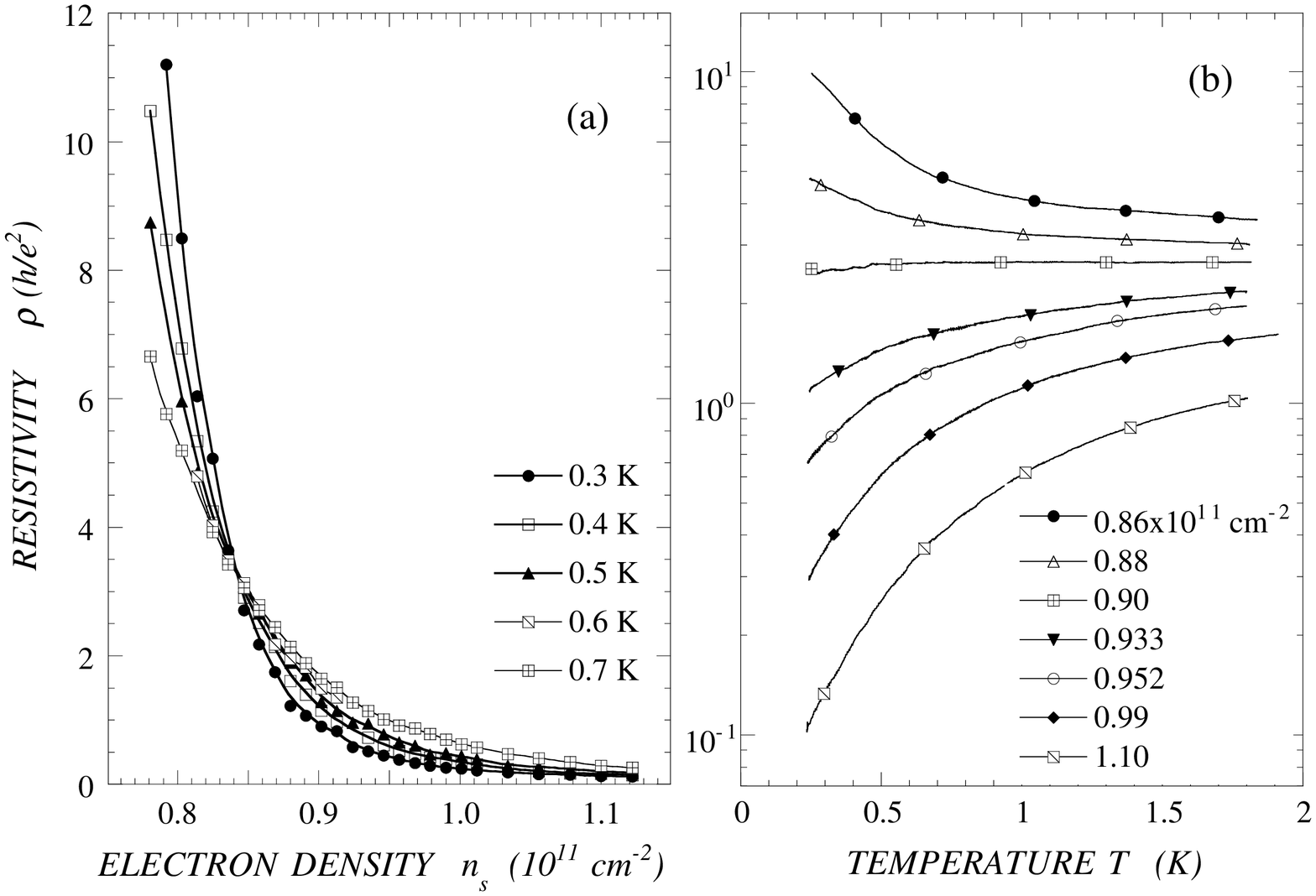} 
}
}
\refstepcounter{figure}
\parbox[b]{7in}{\baselineskip=12pt FIG.~\thefigure.
(a) Resistivity as a function of electron density for the 
two-dimensional system of electrons in a high-mobility silicon 
MOSFET; different curves 
correspond to different temperatures. (b)  Resistivity as a function of 
temperature; here different curves are for different electron densities.
\vspace{0.3in}
}
\label{6}

\vbox{
\vspace{0.2
in}
\hbox{
\hspace{1.5in} 
\epsfxsize 3.5in \epsfbox{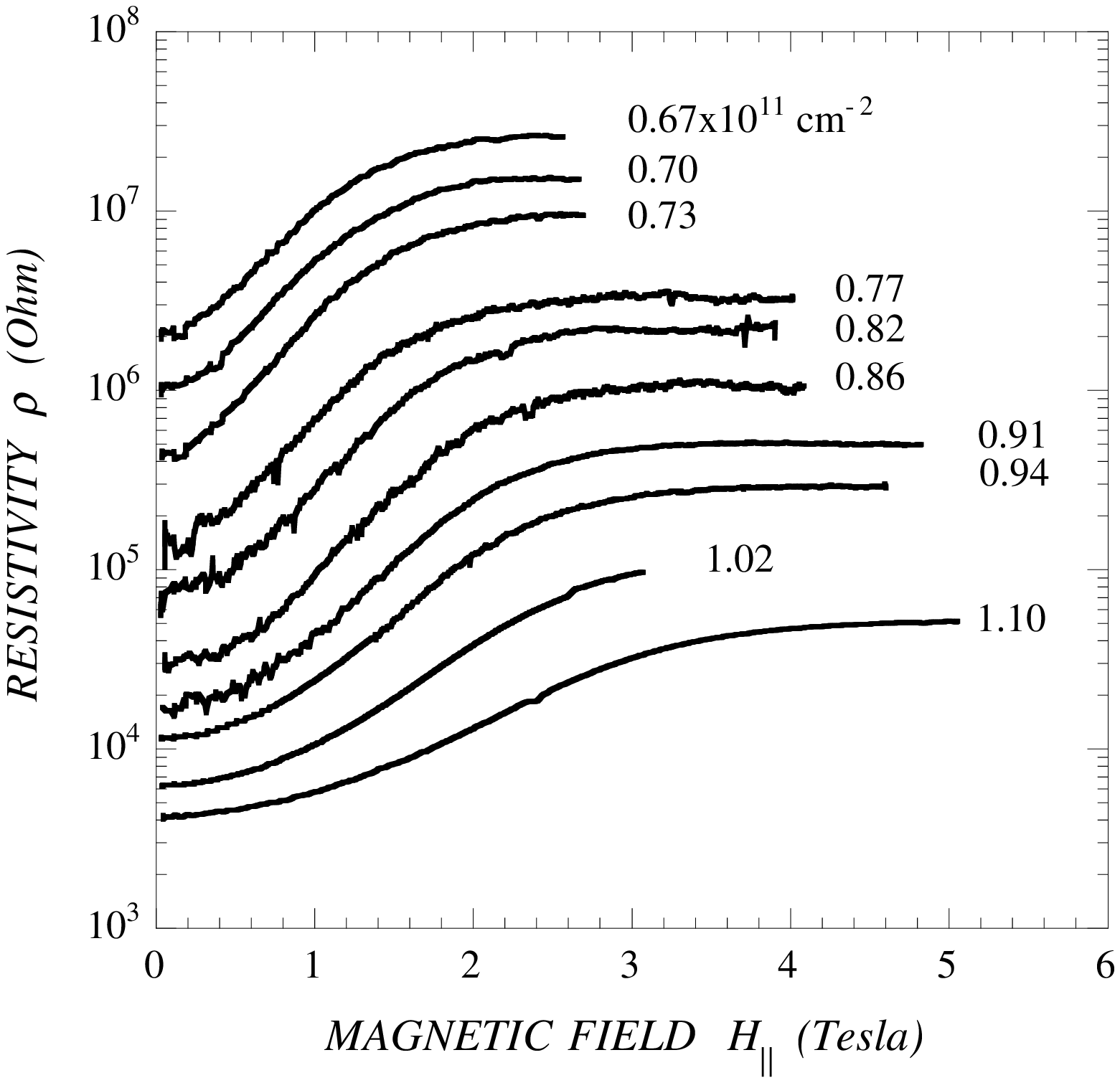} 
}
}
\refstepcounter{figure}
\parbox[b]{7in}{\baselineskip=12pt FIG.~\thefigure.
For different electron densities, the resistivity of a silicon MOSFET at 
0.3 K as a function of magnetic field 
applied parallel to the plane of the 2D electron system.  The top three curves are 
insulating in zero field while the lower curves are conducting.
\vspace{0.3in}
}
\label{7}

A second important feature of these dilute 2D systems is their dramatic response to 
external magnetic fields applied parallel to the plane of the electrons
\cite{simonian}.  As shown in Fig. 7, the resistivity 
increases by several 
orders of magnitude with increasing field and saturates to a constant plateau 
value above a 
density-dependent magnetic field on the order of 2 or 3 Tesla.  The 
magnetoresistance is larger at lower temperatures and in higher mobility samples.  
It should be noted that the curves of Fig. 7 span densities that have both 
metallic and insulating temperature dependence 
in zero field.  The dramatic field-dependence thus appears 
to be a general feature of these dilute two-dimensional electron systems that is 
distinct from the temperature-dependence.

A lively debate has ensued concerning the significance of these findings: whether 
they represent fundamentally new physics or whether they can be explained by an 
extension of physics that is already understood.  A view held by many is that 
these features signal a true zero-temperature quantum phase transition to a novel 
ground state at T=0 ({such as a ``perfect'' metal, a superconductor, a spin 
liquid, a Wigner glass, etc.)
\cite{2Dreview}.   Others argue 
that the anomalous metallic behavior can be explained within a 
single-particle description in terms of a temperature-dependent Drude conductivity, 
and that "metallic" behavior is observed in a restricted range of temperatures so 
that localization prevails in the limit of zero temperature.  Suggestions include 
scattering at charged traps, temperature-dependent screening, interband scattering, 
and thermal smearing of a percolation theshold \cite{2Dreview}.

The high mobilities that are now attainable in MOSFETs and heterostructures 
have opened a new area of investigation in low density 2D electron (hole) systems 
where interactions are very strong.  Regardless of how many-body effects will be 
incorporated into a full description in this regime, the behavior of these materials  
and the physics that is emerging are new and 
fascinating.  My closing remark as an experimentalist is 
that much more experimental information is needed before serious progress can be 
made.

I thank the US Department of Energy for support under grant No. DE-FG02-84-ER45153.


\begin{references}
\bibitem{numbers}  It is important to note that the data of the Bell group shown 
in Fig. 2 and of the Karlsruhe group shown in Fig. 3 differ substantially in 
numerical detail.  This points to a difference between the two experiments associated 
with the samples, the stress, the method of measurement, or some other factor.
\end{references}
\end{document}